\documentclass[aps,physrev,reprint,superscriptaddress]{revtex4-2}

\usepackage{multirow}
\usepackage{amsmath}
\usepackage{graphicx} 
\usepackage{amssymb}
\usepackage{color,xcolor}
\usepackage{textcomp}
\usepackage{booktabs}
\usepackage{rotating} 

\newcommand{\ee}{\mathrm{e}}
\newcommand{\ii}{\mathrm{i}}
\newcommand{\EP}{\text{EP}}
\newcommand{\TP}{\text{TP}}

\begin{document}

\title{5-GHz chip-based quantum key distribution with 1Mbps secure key rate over 150 km}




\author{Guo-Wei Zhang}
    \thanks{These authors contributed equally to this work.}
    \affiliation{Laboratory of Quantum Information, University of Science and Technology of China, Hefei 230026, China}
    \affiliation{Anhui Province Key Laboratory of Quantum Network, University of Science and Technology of China, Hefei 230026, China}
    \affiliation{CAS Center for Excellence in Quantum Information and Quantum Physics, University of Science and Technology of China, Hefei 230026, China}
    
\author{Sheng-Teng Zheng}
    \thanks{These authors contributed equally to this work.}
    \affiliation{Laboratory of Quantum Information, University of Science and Technology of China, Hefei 230026, China}
    \affiliation{Anhui Province Key Laboratory of Quantum Network, University of Science and Technology of China, Hefei 230026, China}
    \affiliation{CAS Center for Excellence in Quantum Information and Quantum Physics, University of Science and Technology of China, Hefei 230026, China}
    \affiliation{Hefei National Laboratory, University of Science and Technology of China, Hefei 230088, China}

\author{You Xiao}
    \thanks{These authors contributed equally to this work.}
   \affiliation{National Key Laboratory of Materials for Integrated Circuits, Shanghai Institute of Microsystem and Information Technology, Chinese Academy of Sciences, Shanghai 200050, China}

\author{Fang-Xiang Wang}
    \email{Corresponding author: fxwung@ustc.edu.cn}
\author{Wen-Jing Ding}
    \affiliation{Laboratory of Quantum Information, University of Science and Technology of China, Hefei 230026, China}
    \affiliation{Anhui Province Key Laboratory of Quantum Network, University of Science and Technology of China, Hefei 230026, China}
    \affiliation{CAS Center for Excellence in Quantum Information and Quantum Physics, University of Science and Technology of China, Hefei 230026, China}
    
\author{Dianpeng Wang}
   \affiliation{National Key Laboratory of Materials for Integrated Circuits, Shanghai Institute of Microsystem and Information Technology, Chinese Academy of Sciences, Shanghai 200050, China}
   
\author{Penglei Hao}
\author{Li Zhang}
\author{Jia-Lin Chen}
    \affiliation{Laboratory of Quantum Information, University of Science and Technology of China, Hefei 230026, China}
    \affiliation{Anhui Province Key Laboratory of Quantum Network, University of Science and Technology of China, Hefei 230026, China}
    \affiliation{CAS Center for Excellence in Quantum Information and Quantum Physics, University of Science and Technology of China, Hefei 230026, China}

\author{Yu-Yang Ding}
\affiliation{Hefei Guizhen Chip Technologies Co., Ltd., Hefei 230000, China}

\author{Shuang Wang}
\author{De-Yong He}
\author{Zhen-Qiang Yin}
    \affiliation{Laboratory of Quantum Information, University of Science and Technology of China, Hefei 230026, China}
    \affiliation{Anhui Province Key Laboratory of Quantum Network, University of Science and Technology of China, Hefei 230026, China}
    \affiliation{CAS Center for Excellence in Quantum Information and Quantum Physics, University of Science and Technology of China, Hefei 230026, China}
    \affiliation{Hefei National Laboratory, University of Science and Technology of China, Hefei 230088, China}

\author{Zheng Zhou}
    \affiliation{Laboratory of Quantum Information, University of Science and Technology of China, Hefei 230026, China}
    \affiliation{Anhui Province Key Laboratory of Quantum Network, University of Science and Technology of China, Hefei 230026, China}
    \affiliation{CAS Center for Excellence in Quantum Information and Quantum Physics, University of Science and Technology of China, Hefei 230026, China}

\author{Hao Li}
    \email{Corresponding author: lihao@mail.sim.ac.cn}
   \affiliation{National Key Laboratory of Materials for Integrated Circuits, Shanghai Institute of Microsystem and Information Technology, Chinese Academy of Sciences, Shanghai 200050, China}
\author{Lixing You}
   \affiliation{National Key Laboratory of Materials for Integrated Circuits, Shanghai Institute of Microsystem and Information Technology, Chinese Academy of Sciences, Shanghai 200050, China}

\author{Guang-Can Guo}
\author{Wei Chen}
    \email{Corresponding author: weich@ustc.edu.cn}  
\author{Zheng-Fu Han}
  \affiliation{Laboratory of Quantum Information, University of Science and Technology of China, Hefei 230026, China}
    \affiliation{Anhui Province Key Laboratory of Quantum Network, University of Science and Technology of China, Hefei 230026, China}
    \affiliation{CAS Center for Excellence in Quantum Information and Quantum Physics, University of Science and Technology of China, Hefei 230026, China}
    \affiliation{Hefei National Laboratory, University of Science and Technology of China, Hefei 230088, China}




\begin{abstract}
Quantum key distribution (QKD) enables secure communication by harnessing the fundamental principles of quantum physics, which inherently guarantee information-theoretic security and intrinsic resistance to quantum computing attacks. However, the secure key rate of QKD typically decreases exponentially with increasing channel distance. In this work, by developing a novel polarization-state preparation method, an ultra-low time-jitter laser source and superconducting nanowire single-photon detectors, we demonstrate a 5-GHz integrated QKD system featuring ultra-low quantum bit error rates (QBERs). The system achieves secure key rates of 1.076 Mbps at 150 km and 105 kbps at 200 km over standard single-mode fiber channels, respectively. Our system substantially enhances the secure key rate, enabling high-resolution video calls with one-time-pad encryption over intercity backbone QKD links. This work represents a significant step forward in the development of high-performance practical QKD systems.
\end{abstract}



\maketitle

\section{Introduction}
\label{section:introduction}

Quantum key distribution (QKD) encodes the key into quantum states so that eavesdropping across the quantum channel can be perceived by monitoring the quantum bit error rates (QBERs) of photon states \cite{Bennett1984_ICCSSP_QCPK, Ekert1991_PRL_QCBB,wangxb2005_decoy,lo2005_decoy}. Owning to its high security level, QKD has been deployed from metropolitan to intercity and integrated space-to-ground quantum networks \cite{Frohlich2013_BB84_network,wangs2014_bb84_network,chenya2021_space-to-ground_network}. Secure key rate (SKR) is the core indicator for practical applications of QKD. However, the single-photon-level quantum states are attenuated exponentially across the quantum channel, so does the SKR \cite{pirandola2017_plob}. 
Increasing the operating rate of the QKD system from MHz to GHz is the simplest yet most effective approach to raise the SKR \cite{takesue2006_10ghz,boaron2018a_2.5ghz,wangshuang2018_1ghz,yuan2018_13mbps_1ghz,grunenfelder2020_5ghz}. However, as the operating rate increases, maintaining relatively high signal-to-noise ratios (SNRs) during the preparation and detection procedures becomes a critical challenge, which limits the enhancement to SKR \cite{wangshuang2018_1ghz,grunenfelder2020_5ghz}. 

Integrated photonics, leveraging inherent advantages such as compact footprint, scalability, high modulation speed, low manufacturing cost, and high-volume production feasibility, is a promising platform for practical deployment of QKD \cite{sibson2017_nc_on-chip,sibson2017_on-chip,wei2020_prx_mdi_on-chip,paraiso2021_np_on-chip,zhanggw2022_plc_on-chip}. Benefiting from integrated encoding chips, gigahertz QKD system have garnered significant research advances \cite{beutel2022_fifw_on-chip,sax2023_2.5ghz_on-chip,Lin2025_LN_on-chip} and achieved Mbps SKR over 100 km fiber channel \cite{liw2023_110mbps_on-chip}. Nevertheless, to increase SKR further and to extend the Mbps SKR to longer distance are still the persistent pursuits for large-scale intercity quantum networks \cite{chenya2021_space-to-ground_network}.

However, increasing the operating rate does not deservedly guarantee a higher SKR. A higher operating rate leads to degraded state preparation fidelity and more significant time jitter errors from the photon sources and single-photon detection process \cite{Grunenfelder2023_64Mbps_10km}. Otherwise, a higher operating rate may lead to a higher QBER and lower SKR \cite{grunenfelder2020_5ghz} than that with lower operating rate \cite{sax2023_2.5ghz_on-chip,liw2023_110mbps_on-chip}.
Therefore, A high-SKR high-speed QKD system requires simultaneous optimization of state preparation fidelity and time jitters of light source and single-photon detectors, which is still to be addressed for higher than 2.5 GHz system.

In this work, we significantly improved the high-speed performance of key modules, including the narrow temporal-width laser, ultra-low time jitter superconducting nanowire single-photon detector (SNSPD) chips and their high SNR read-out circuits and a high-performance integrated QKD transmitter chip. Based on these technique advances and a high-precision polarization state preparation method, we successfully develop an integrated 5-GHz QKD system with a QBERs suppressed to about 0.5\%. Ultimately, we achieve a SKR exceeding 1 Mbps over a 150 km standard single-mode fiber channel, which is one order of magnitude higher than previous state-of-the-art QKD systems. The system shows a long-term stability with continuous running. Our work strongly demonstrates the feasibility of a practical, high-performance integrated QKD system capable of enabling high SKR between intercity backbone quantum links.

\section{QKD system and states preparation}
\label{section:setup}

\subsection{QKD system}
\label{subsection:setup}

\begin{figure*}[htbp]
	\centering
	\includegraphics[width=0.9\textwidth]{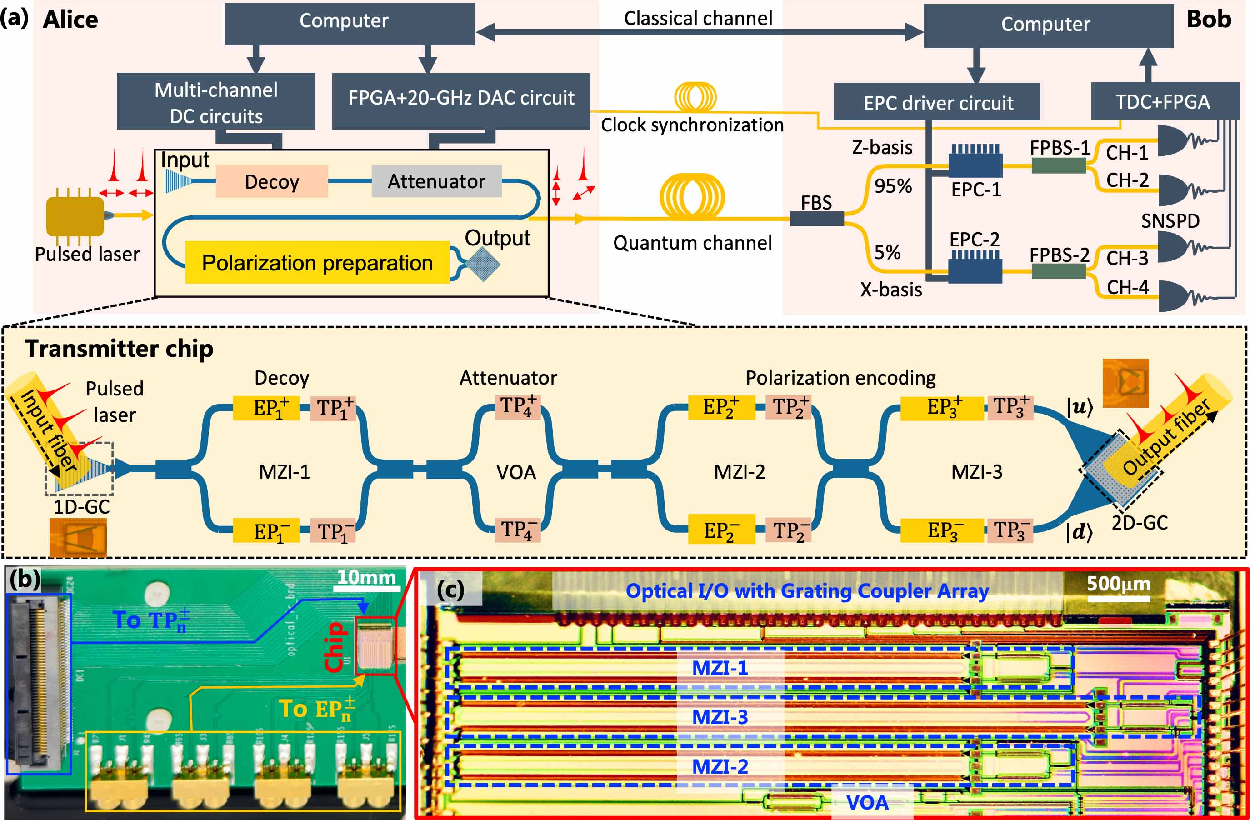}
	\caption{QKD system with integrated transmitter chip. (a) Schematic setup of the QKD system and the transmitter chip architecture. The chip consists of three high-speed Mach-Zehnder modulators (MZI-1 to MZI-3) and one thermal Mach-Zehnder modulator(VOA). MZI-1 is employed for decoy state preparation;  MZI-2 and MZI-3 are for polarization encoding. VOA is used as an attenuator. The superscript $\pm$ inside the phase modulator indicates the upper and lower paths of the Mach-Zehnder modulator, respectively. (b) The opto-electronic packaging of the integrated photonic chip with compact printed circuit board, and (c) the microscope image of the chip corresponding to the red-boxed area.}
	\label{fig:setup}
\end{figure*}

The QKD system employs a polarization-encoding BB84 protocol using one-decoy state method \cite{rusca2018_onedecoy_skr} with two unbiased basis. The Z basis ($|H\rangle$ and $|V\rangle$) is chosen randomly with higher probability for secret key extraction, while X basis ($|\pm\rangle=\frac{1}{\sqrt{2}}(|H\rangle\pm e^{\ii\chi}|V\rangle)$) is used for eavesdropping detection, where $|H\rangle$ and $|V\rangle$ are the horizontal and vertical polarizations, respectively, and $\chi$ is an arbitrary initial phase. The QKD system is shown in Fig. \ref{fig:setup}(a). For Alice, the distributed feedback (DFB) laser is incident into the silicon-based transmitter chip through a one-dimensional grating coupler (1D-GC). The silicon-based transmitter chip is controlled and driven by a custom-designed circuit module, which comprises a field-programmable gate array (FPGA), a 64-channel constant current and constant voltage source, and a high-speed digital-to-analog converter (DAC), enabling precise generation of different polarization states.  

Within the chip, carrier-depletion electro-optic modulators (EPs, yellow rectangles) and thermo-optic modulators (TPs, pink rectangles) are used to realize high-speed electrical and broad-range thermal phase modulations, respectively. High-speed Mach–Zehnder interferometers (MZIs) are utilized to modulate decoy intensities (MZI-1) and to encode polarization states (MZI-2 and MZI-3). A thermally tuned MZI functions as a variable optical attenuator (VOA) to attenuate the pulses into the single-photon level. The two-dimensional grating coupler (2D-GC) maps the two paths and their superpositions of MZI-3 into polarization states (Z and X bases). The transmitter chip is electronically packaged on a compact printed circuit board ($\rm 115\times 38\ mm^2$, Fig. \ref{fig:setup}(b)). The optical input and output facets of the photonic chip are coupled to a fiber array (see the micro-image in Fig. \ref{fig:setup}(c)).

The QKD receiver passively chooses Z or X basis using an unbalanced fiber beam splitter (FBS). Polarization drifts in the fiber channel is compensated for by two electro-optic polarization controllers (EPCs). Photons are projected and measured via fiber polarized beam splitters (FPBSs) paired with a four-channel SNSPD. Detection results are recorded and analyzed by a time-to-digital converter (TDC) and FPGA. Alice and Bob control the system using computers and communicate through a classical channel, while clock synchronization is achieved via another independent fiber channel.

\subsection{Performances of key modules}
\label{subsection:Laser_SNSPD_Chip}

\begin{figure*}[htbp]
	\centering
	\includegraphics[width=0.9\textwidth]{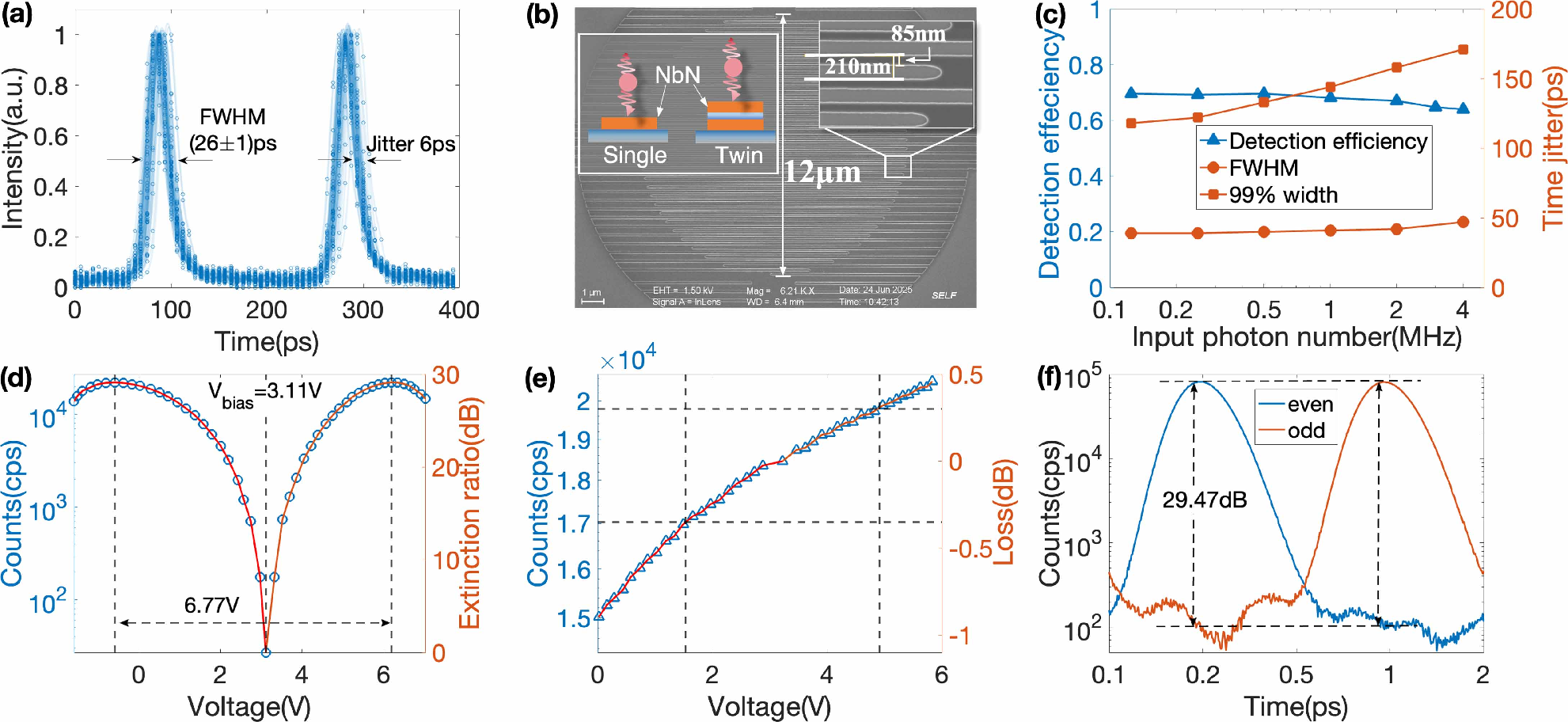}
	\caption{Laser source, SNSPD and transmitter chip. (a) The temporal profiles of the gain-switched  laser with 100 measurements. (b) The scanning electron microscope (SEM) image of the twin-layer SNSPD chip. (c) Detection efficiency (blue, left axis) and time jitters (red, right axis) of the SNSPD with different input photon intensities. (d) Voltage-dependent transmission of the MZI and (e) loss of the carrier-depletion phase modulator. (f) The ERs of the MZI to the 5-GHz pulses by suppressing the even (blue)/odd (red) pulses.}
	\label{fig:laser_SNSPD_MZI}
\end{figure*}

The performance of a high-speed QKD system is primarily determined by three parameters: 1) the temporal width of the laser pulse, 2) the time jitter of the single photon detectors and 3) the extinction ratio (ER) of the encoding scheme, including its implementation. For a 5-GHz QKD system, the combination of higher bandwidth demands and the very short 200-ps interval between adjacent pulses imposes stricter requirements on system design and implementation.  


The DFB laser operates in gain-switching mode and effectively suppresses its temporal pulse width by integrating multiple techniques. These techniques consist high-precision radio frequency (RF) modulation, fine bias tuning, and stable temperature control. Fig. \ref{fig:laser_SNSPD_MZI}(a) gives the temporal profile of the pulsed laser with 100 measurements. The statistical analysis shows that the full width at half maximum (FWHM) and time jitter (root mean square) of the laser pulse are $(26\pm1)$ ps and 6 ps, respectively. It is one order of magnitude smaller than 200 ps and significantly outperforms the previous 5-GHz QKD system\cite{grunenfelder2020_5ghz}.  

As high-speed QKD system requires ultra-low detection time jitter, it leads to low nanowire duty cycle, which would decrease the detection efficiency of the SNSPD. We design and fabricate a high-performance SNSPD chip architecture with twin-layer-NbN nanowires to balance the performance between detection efficiency and time jitter \cite{Hu2020_OE_DSIP}. Figure \ref{fig:laser_SNSPD_MZI}(b) presents the (enlarged) image of the fabricated nanowire chip taken with a scanning electron microscope (SEM). The diameter of the sensitive area of the SNSPD chip is about 12 \textmu m. The width and duty cycle of the nanowire are 85 nm and 40.5\%, respectively. As shown in Fig. \ref{fig:laser_SNSPD_MZI}(c), the detection efficiency only slightly decreases from 70\% to 64\% as input photon intensity increases from 125 kHz to 4 MHz (left y axis, blue triangles). Meanwhile, the FWHM (red circles) and full width at 1\% maximum (red squares) of the detection time jitter are suppressed to be less than 50 ps and 171 ps, respectively. Hence, the twin-layer SNSPD chip possesses high detection efficiency, high count rate and low time jitter simultaneously and is competent for a high-performance QKD system.

The element units on the transmitter chip are the MZIs. The MZIs of the transmitter chip are operated in the push-pull modulation mode. The voltage-dependence of the MZI's transmittance is illustrated in Fig. \ref{fig:laser_SNSPD_MZI}(d). The 2$\pi$ voltage of the MZI is about $\rm V_{2\pi}=6.77$ V and the intensity ER is close to 30 dB (red line). The voltage-dependent insertion loss of the EP is about 0.19 dB/V (Fig. \ref{fig:laser_SNSPD_MZI}(e)), contributing to an intrinsic optical error rate of less than 0.14\%. When driven by a 5-Gbps square electrical signal, the MZI suppresses the odd/even positions of the 5-GHz laser pulse sequence with an ER of about 29.47 dB (Fig. \ref{fig:laser_SNSPD_MZI}(f)). 

\begin{figure*}[htbp]
	\centering
	\includegraphics[width=\textwidth]{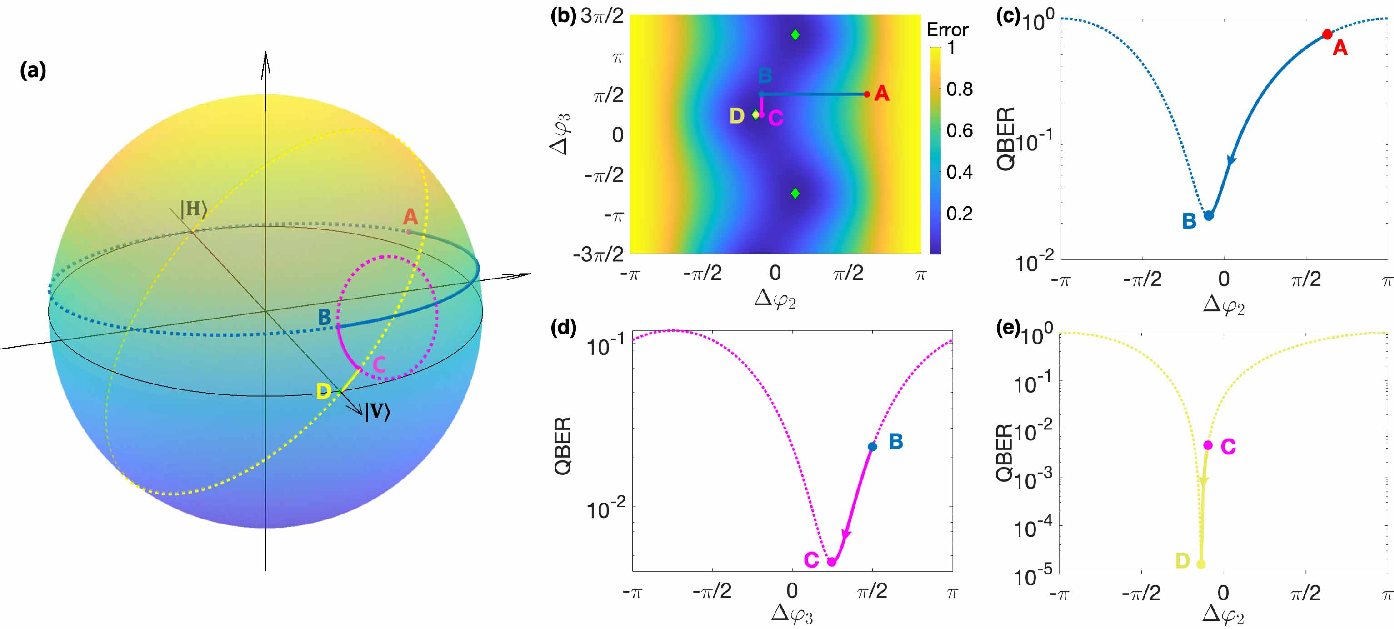}
	\caption{Concept diagram for perfect polarization state preparation. (a) The iterative trajectory curves of imperfect $|V'\rangle$ to the ideal $|V\rangle$ on the Poincar\'e sphere. (b) The error rate $R_{err}$ with respect to $\Delta\varphi_2$ and $\Delta\varphi_3$. A three-step iterative process is enough to approach a negligible error. (c)-(e) The iterative trajectory curves of $R_{err}$ from the original point A to the nearly perfect point D.}
	\label{fig:state_preparation_methods}
\end{figure*}

\subsection{States preparation}
\label{subsection:StatesPreparation}

Due to the requirement of precise nano-structure to realize high ER, the 2D-GC demands the 10-nm-level manufacturing precision in multiple factors, such as hole size and etch depth. Therefore, it exhibits sensitive to microscopic parameters and wavelength \cite{Sobu2018_JJAP_STGC}. Furthermore, when coupled with a 2D-GC, the fiber requires tilted incidence to provide directionality and reduce back-reflection. This breaks the ideal symmetry and thereby introducing significant polarization sensitivity. Thus, achieving a polarization ER of >23 dB, which is high enough to meet the requirements of practical QKD chips, remains a challenging task. That is to say, even though the high ER MZI guarantees the path chosen purity of $|u\rangle$ and $|d\rangle$, the combined effects of this polarization sensitivity and finite manufacturing tolerances still result in limited polarization ERs and non-negligible QBERs \cite{Carroll2013_OE_BPOP}. The polarization isolation of the 2D-GC in this work is only 15.8 dB, which may lead to a non-negligible intrinsic QBER ($\sim2.63\%$). 

Here, by utilizing a dual-MZI structure (MZI-2 and MZI-3 in Fig. \ref{fig:setup}(a)), we propose a novel polarization states preparation method that is able to prepare two sets of nearly perfect mutually unbiased bases (MUBs) in polarization even with a low-ER (e.g., 10-dB) 2D-GC. It releases the stringent manufacturing technique requirements on the photonic chip. 

Considering the imperfections, the transformation matrix of the 2D-GC becomes $M_{\text{2D-GC}}=|H\rangle\langle u|+\ee^{\ii\zeta}\sin\delta |H\rangle\langle d| + \cos\delta |V\rangle\langle d|$, where $\delta$ and $\zeta$ are determined by the 2D-GC and can be measured in experiment. The output polarization state after passing through the 2D-GC is given by: 
\begin{widetext}
    \begin{equation}
        \begin{aligned}            |\psi\rangle_{out}=\left[\ee^{\ii\Delta\varphi_3}t_{\EP_3^+}\sin\frac{\Delta\varphi_2}{2} + \ee^{\ii\zeta}t_{\EP_3^-}\sin\delta\cos\frac{\Delta\varphi_2}{2}\right]|H\rangle+t_{\EP_3^-}\cos\delta\cos\frac{\Delta\varphi_2}{2}|V\rangle
        \end{aligned}
        \label{eq:outputpolarization}
    \end{equation}
\end{widetext}  
where $\Delta\varphi_n = \Delta\varphi_{\EP_n} + \Delta\varphi_{\TP_n}$ is the phase difference between the two paths of MZI-$n$, resulting from the combination of the phase differences in the EPs ($\Delta\varphi_{\EP_n}$) and TPs ($\Delta\varphi_{\TP_n}$). $t_{\EP_3^\pm}$ are the transmission coefficients of $\EP^{\pm}_3$, with superscripts $\pm$ corresponding to the upper and lower path, respectively. According to Eq. (\ref{eq:outputpolarization}), let $\Delta\varphi_2=0$ and the photon goes through the upper path of MZI-3 only, then the final output polarization becomes $|H\rangle$. For $|V\rangle$ state, without loss of generality, we assume the initial dual-MZI parameters begins the imperfect vertical polarization $|V'\rangle$ from point \textbf{A} on the Poincar\'e sphere of Fig. \ref{fig:state_preparation_methods}(a). By iteratively tuning the phases of $\EP^{\pm}_{2}$ and $\TP^{\pm}_{3}$, $|V'\rangle$ can be rotated to the ideal $|V\rangle$ (point \textbf{D}) rapidly in a convergent way (the solid trajectory curves \textbf{A}$\xrightarrow{}$\textbf{B}$\xrightarrow{}$\textbf{C}$\xrightarrow{}$\textbf{D}). Figure \ref{fig:state_preparation_methods}(b) shows the error rate of $|V'\rangle$ from $|V\rangle$ ($R_{err}=\big|\langle H|V'\rangle\big|^2$) with respect to $\Delta\varphi_2$ and $\Delta\varphi_3$. As proved in Appendix \textbf{\ref{appendix:Z_basis}}, the solution to $R_{err}=0$ exists (the greed diamonds). Furthermore, $R_{err}$ is monotonic with respect to $\Delta\varphi_2$ and $\Delta\varphi_3$ within $(-\pi/2,\pi/2]$. As shown in Figs. \ref{fig:state_preparation_methods}(c)-\ref{fig:state_preparation_methods}(e), the iteration process could suppress $R_{err}$ from 0.7374 to below $10^{-4}$ within 3 steps.

For a perfect X basis, it should satisfies the form $|\pm\rangle=\frac{1}{2}(|H\rangle\pm \ee^{i\chi}|V\rangle)$. As $\Delta\varphi_{\TP_2}$ and $\Delta\varphi_{\TP_3}$ have been fixed to guarantee the preparation of Z basis, the preparation of X basis relies on $\EP^{\pm}_2$ and $\EP^{\pm}_3$. Similarly, it can be proved that a perfect X basis can be prepared by iteratively modulating $\Delta\varphi_{\TP_2}$ and $\Delta\varphi_{\TP_3}$. Thanks to the monotonicity, the iterative process can be completed quickly (see Appendix \textbf{\ref{appendix:xbasis}} for more details). As given by TABLE \ref{tab:sim_results}, the proposed method here can realize zero-QBER polarization states preparation in principle with negligible intensity dependence, even when the ER of the 2D-GC is as low as 10 dB (see Appendix \textbf{\ref{appendix:state-dependet intensity}} for more details). It means very small state-dependent intensities and has negligible effect on secret key extraction \cite{StateDependentIntensity,liw2023_110mbps_on-chip}. This significantly enhance the performance of transmitter chips under realistic fabrication tolerances.

\section{Results}
\label{section:results}

\begin{figure*}[htbp]
	\centering
	\includegraphics[width=\textwidth]{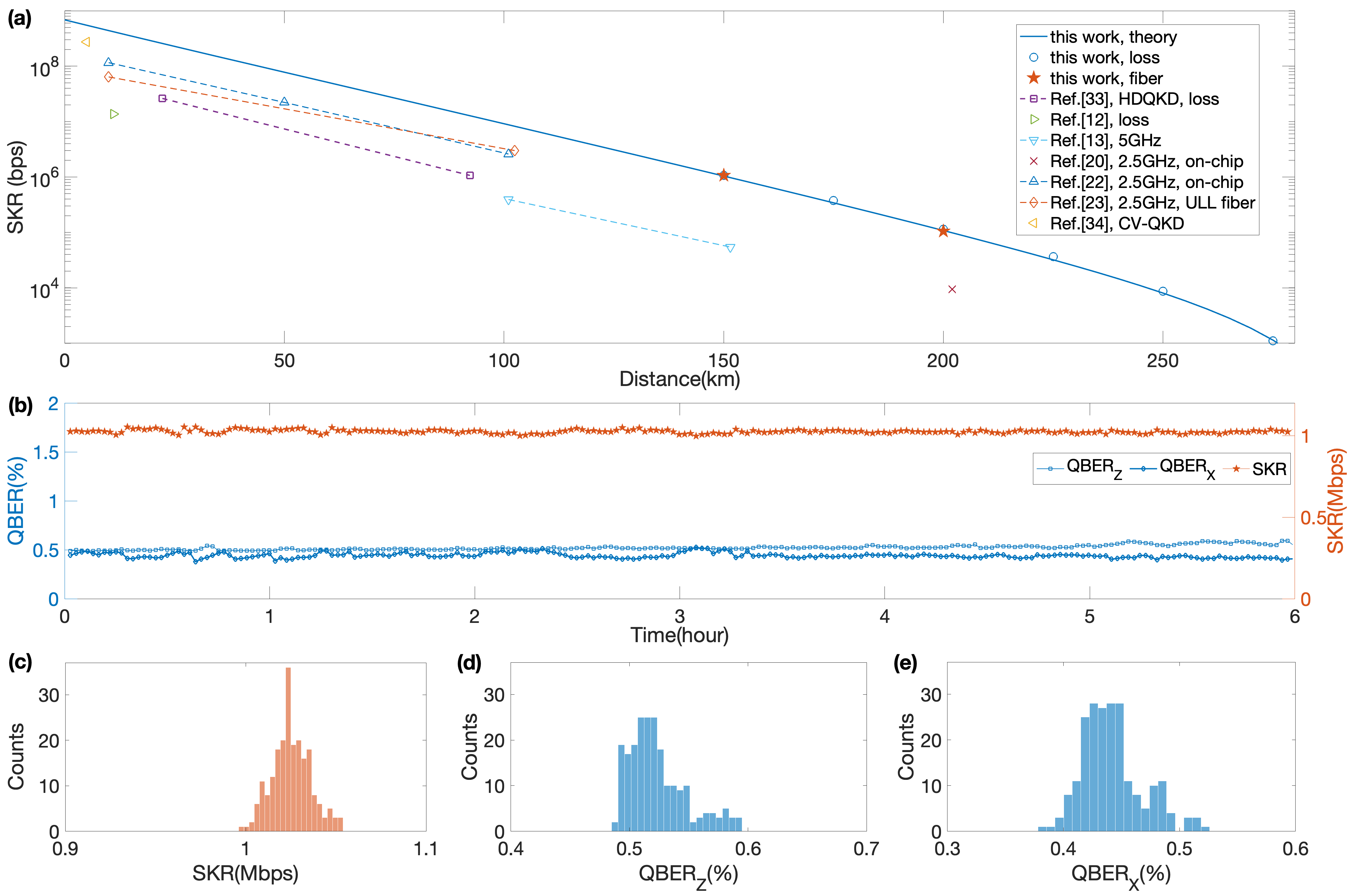}
	\caption{SKR and stability of the QKD system. (a) SKR over distance. (b) Real-time SKR and QBERs for 6 hours. (c)-(e) Frequency statistics on SKR, QBERs of Z and X bases, respectively.}
	\label{fig:figureSKR}
\end{figure*}

Based on the high-performance laser source, transmitter chip, SNSPD and state preparation method, a high-performance 5-GHz QKD system is realized. The QKD system employs the BB84 protocol \cite{Bennett1984_ICCSSP_QCPK} and one-decoy-state method \cite{wangxb2005_decoy,lo2005_decoy,rusca2018_onedecoy_skr}. Considering the finite-key-length effect \cite{lim2014_finite_size_effect_skr}, the secure keys are extracted every 100 seconds according to the following formula \cite{rusca2018_onedecoy_skr}
\begin{equation}
    \centering
	l\leq s_\text{Z,0}^l+s_\text{Z,1}^l(1-h(\phi_\text{Z}^u))-\lambda_\text{EC}-6\log_2(\frac{19}{\epsilon_\text{sec}})-\log_2(\frac{2}{\epsilon_\text{cor}})
 	\label{eq:secure_key_rate}
\end{equation} 
where $s_\text{Z,0}$ and $s_\text{Z,1}$ are the lower bound on the vacuum and single-photon events, respectively, $\phi_\text{Z}^u$ is the estimated upper bound on the phase error rate of Z basis, $\lambda_\text{EC}$ is the key consumption during post-processing procedure, and $\epsilon_\text{sec}$ and $\epsilon_\text{cor}$ are the secrecy and correctness parameters (see Appendix \textbf{\ref{appendix:securekeyrate}} for details).

The system is demonstrated over 150-km and 200-km fiber channels (standard single-mode fiber,  ITU-T G.652.D, 0.18dB/km) and equivalent loss channels from 150 km to 275 km. Polarization drifts in the fiber channel are compensated by the EPCs using the stochastic parallel gradient descent (SPGD) algorithm. Owing to the proposed dual-MZI polarization states preparation method, the QBER of the system with a 150-km (200-km) fiber is as low as 0.47\% (0.51\%), which is very close to the 0.38\% (0.31\%) QBER of the equivalent-loss channel, respectively. The SKR of our QKD system is demonstrated in Fig. \ref{fig:figureSKR}(a), where the SKRs over 150-km and 200-km fiber are 1.076 Mbps and 105 kbps, respectively. This enables real-time video and voice calls between long-distance intercity users using one-time pad encryption. 

As presented in Fig. \ref{fig:figureSKR}(b), the implemented chip-based QKD system is stable for long-time running. The QBERs (blue circles and diamonds) and SKR (red stars) remain highly stable over 6 hours' continuous running with a 150-km-fiber channel. Statistical analysis reveals an average SKR of $1.025\pm0.011$ Mbps (Fig. \ref{fig:figureSKR}(c)), with QBERs of Z and X bases are $(0.52\pm0.02)\%$ and $(0.44\pm0.03)\%$, respectively. These results conclusively demonstrate the high performance and exceptional stability of our QKD system. 

\section{Discussion and conclusion}
\label{section:conclusion}

We have proposed an analytical state preparation method that releases the ultra-high polarization ER requirement to the 2D-GC and experimentally realize ultra-low-QBER MUBs. By preparing the Y basis(left- and right-circular polarization) instead of the Z basis, the negative impact of the low-ER 2D-GC may be mitigated to a certain extent. However, such a method requires a larger modulation voltage of $3\pi/2$ for the EPs and deteriorates the ER of the MZI \cite{wei2020_prx_mdi_on-chip}. It also demands complex pre-calibration and leads to polarization-dependent ER \cite{liw2023_110mbps_on-chip}. While our method overcomes the disadvantages above and. We also design and fabricate a twin-layer architecture SNSPD chip to achieve high detection efficiency and low time jitter simultaneously, which significantly suppresses the temporal crosstalk between time bins separated by only 200 ps. Thanks to the state preparation method proposed and high-quality integrated techniques advanced in our system, we finally realize the first 5-GHz integrated QKD system and extend the distance for Mbps-level secure-key-rates (supporting real-time video calls) from 100 km \cite{liw2023_110mbps_on-chip} to 150 km, which is 20 times higher than previous fiber-based 5-GHz QKD system \cite{grunenfelder2020_5ghz} (the blue inverted triangles in Fig. \ref{fig:figureSKR}(a)). The SKR of our system over 200 km is also one order of magnitude larger than those of state-of-the-art systems \cite{grunenfelder2020_5ghz,sax2023_2.5ghz_on-chip}. 

Our system also demonstrates that the BB84 protocol achieves a higher SKR within 200 km than other long-distance protocols such as twin-field QKD, which exhibits significant advantages for ultra-long-distance communication \cite{wangs2022_830km_tf,liuy2023_1000km_tf,du2024_o_tfqkd_on-chip}. Furthermore, theoretical simulation indicates that our system has the potential to reach a SKR of 550 Mbps at 5 km (228 Mbps@25 km) with a single channel. It is significantly higher than the start-of-the-art high-dimensional QKD \cite{Islam2017a_HDQKD_4D} and continuous-variable (CV) QKD \cite{ji2024_cvqkd}, both of which are promising protocols for high-speed quantum access networks in the metropolitan areas. 

In conclusion, we have realized the first integrated QKD system operating at 5 GHz. The QKD chip is opto-electronic packaged with high bandwidth, and the ER of the on-chip MZI approaches 30 dB. The temporal width of the pulsed laser is as less as 26 ps. By developing a precise state-preparation method, we achieve the generation of polarization states with ultra-low-QBER using a 2D-GC featuring a relatively low ER. Furthermore, we have developed a high-performance SNSPD chip, where its twin-layer architecture with low duty cycle ensures both high detection efficiency and ultra-low detection time jitters simultaneously. Finally, we have achieved the first Mbps SKR QKD system over a 150-km fiber channel, which is one to two orders of magnitude larger than that of state-of-the-art implemented QKD systems. Our work demonstrates that the BB84 protocol has the potential to achieve the highest SKR from metropolitan to intercity QKD networks within 200 km and significantly improves the performance of practical QKD systems.

\section*{Acknowledgments}
This work has been supported by National Natural Science Foundation of China (Grant No. 62371437), Innovation Program for Quantum Science and Technology (Grant No. 2021ZD0300701), and Industrial Prospect and Key Core Technology Projects of Jiangsu Provincial Key R \& D Program (BE2022071).
This work was partially carried out at the USTC Center for Micro and Nanoscale Research and Fabrication.

\section*{Competing interests}
The authors declare that they have no competing interests.
\section*{Data and materials availability}
All data needed to evaluate the conclusions in the paper are present in the paper and the Appendix.



\bibliography{5GRef}
\end{document}